\documentclass[aps,notitlepage,a4paper,onecolumn,showpacs,showkeys,floatfix,superscriptaddress]{revtex4-1}

\newcommand{\AlMn}{Mn$_3$Al$_2$Ge$_3$O$_{12}$}
\newcommand{\csnicl}{CsNiCl$_3$}
\newcommand{\rbnicl}{RbNiCl$_3$}
\newcommand{\licuo}{LiCu$_2$O$_2$}
\newcommand{\gto}{Gd$_2$Ti$_2$O$_7$}
\newcommand{\sumprimed}[1]{\sideset{}{'}\sum_{#1}}
\newcommand{\figwidth}{0.98\columnwidth}

\usepackage[dvips]{epsfig}
\usepackage{amsmath}

\begin{document}

\title{Numeric calculation of antiferromagnetic resonance frequencies for the noncollinear antiferromagnet}

\author{V.Glazkov}
\email{glazkov@kapitza.ras.ru}
\homepage{http://www.kapitza.ras.ru/rgroups/esrgroup/}
\affiliation{P.Kapitza Institute for Physical Problems RAS, Kosygin str.2, 119334 Moscow, Russia}
\affiliation{Moscow Institute of Physics and Technology, 141700 Dolgoprudny, Russia}

\author{T.Soldatov}
\affiliation{P.Kapitza Institute for Physical Problems RAS, Kosygin str.2, 119334 Moscow, Russia}
\affiliation{Moscow Institute of Physics and Technology, 141700 Dolgoprudny, Russia}
\author{Yu.Krasnikova}
\affiliation{P.Kapitza Institute for Physical Problems RAS, Kosygin str.2, 119334 Moscow, Russia}
\affiliation{Moscow Institute of Physics and Technology, 141700 Dolgoprudny, Russia}

\date{\today}

\begin{abstract}
We present an algorithm for the numeric calculation of antiferromagnetic resonance frequencies for the non-collinear antiferromagnets of general type. This algorithm uses general exchange symmetry approach \cite{andrmar} and is applicable for description of low-energy dynamics of an arbitrary noncollinear spin structure in weak fields. Algorithm is implemented as a MatLab and C++ program codes, which are available for download. Program codes are tested against some representative analytically solvable cases.
\end{abstract}

\maketitle

\tableofcontents

\part{Main matter}
\section{Introduction}
Antiferromagnetic ordering is observed at low temperatures in a vast amount of crystals. In the simplest case magnetic ions can be grouped in two sublattices with antiparallel average spins yielding the collinear antiferromagnetic structure. However, such collinear structures do not cover all possible types of antiferromagnetic order: one can easily imagine magnets with more then two sublattices or a helicoidal structures, which can not be deduced to a finite number of sublattices at all. Numerous examples of such systems are known, e.g.: three-sublattice ``triangular'' antiferromagnetic order in \csnicl{} and \rbnicl{}\cite{csnicl-neutrons,csnicl-afmr}, 12-sublattices ordering in an \AlMn{} garnet \cite{garnet-prandl,garnet-afmr}, spiral ordering in \licuo{} \cite{licuo-nmr,licuo-neutrons,svistov-licuo}, complicated multi-$\vec{k}$ structure in strongly frustrated pyrochlore magnet \gto{} \cite{stewart}.

Electron spin resonance (antiferromagnetic resonance, AFMR) is a powerful and sensitive tool to study low-energy dynamics of the magnetically ordered systems. By exciting uniform oscillations (i.e. $k=0$ spin waves) of the ordered  spin structure one can investigate its properties: orientation of the ordered structure with respect to the crystal, strength of the anisotropic interactions fixing this orientation, various spin-reorientation transitions. Due to high energy resolution of microwave AFMR spectrometry spin waves spectrum details at $\vec{k}=0$ can be quite routinely determined with accuracy up to 5 mkeV (corresponding to the resolution of 1 GHz), thus strongly complimenting powerful inelastic magnetic neutron scattering techniques at low energies.

However, interpretation of the antiferromagnetic resonance data for complicated magnetic structures is sometimes difficult. Antiferromagnetic resonance frequencies for a collinear structure can be calculated relatively easily in a two-sublattice model \cite{nagomiya}. Similar calculations for a noncollinear magnets are much less general: many-sublattice model calculations using a mean-field theory approach are very cumbersome \cite{tanaka}, standard (Holstein-Primakoff like) spin-wave theory approach to many-sublattice antiferromagnets is also complicated (e.g., \cite{chubukov,sosin2009}). Moreover, analytical solution of these equations is usually out of the question for general mutual orientation of the magnetic field and sublattices magnetizations. Numeric calculations of spin waves spectra are also known, see e.g. SpinW library by S.T\'{o}th \cite{spinW}, but they rely on strongly model dependent microscopic hamiltonian.

Some of these difficulties can be overcome by using an exchange symmetry approach developed in \cite{andrmar}. This approach allows to build up general hydrodynamic description of low-energy dynamics of an antiferromagnet. It was successfully applied for various magnetic systems \cite{csnicl-afmr,garnet-afmr,svistov-licuo,sosin2006,mar1999,vasiliev2001}. However, analytical solution for $f(H)$ dependency (which is the characteristic observable in antiferromagnetic resonance experiment) remains complicated, if possible at all, for arbitrary direction of magnetic field.

In the present manuscript we describe numeric approach to the solution of dynamic equations in the arbitrary case. The developed algorithm is implemented in a program codes, available free of charge on the authors web-page (http://www.kapitza.ras.ru/rgroups/esrgroup/) \cite{home}.

\section{Brief basics of exchange symmetry approach and derivation of dynamics equations}
First, we briefly recall necessary equations of exchange symmetry theory \cite{andrmar} to be used in our calculations. The main limitation of this theory is that distortions of the ordered spin structure are small, which limits its applicability to the low fields $H\ll H_{ex} \simeq J/(g\mu_B)$. In particular, this limitation excludes from consideration various phase transitions with complete restructuring of the order parameter (collinear-noncollinear transitions, various magnetization plateau phases etc.). Under this assumption any noncollinear magnetic structure can be described by three unitary orthogonal vectors $\vec{l}_{1,2,3}$ (e.g., planar structure with a wavevector $\vec{k}$ can be described as $\vec{S}(\vec r)={\vec l}_1 \cos(\vec k \vec r)+{\vec l}_2 \sin(\vec k \vec r)$, with ${\vec l}_3={\vec l}_1\times{\vec l}_2$). All static  properties and low energy dynamic of this structure can be described by its Lagrangian with Lagrangian density (we use here notations of Refs. \cite{svistov-licuo,far-glaz})

\begin{equation}
{\cal L}=\sum_i \frac{I_i}{2}  \left(\dot{\vec{l}}_i +\gamma \left[\vec{ l}_i \times
  \vec{ H} \right] \right)^2 - U_A(\left\lbrace {\vec l}_i\right \rbrace)
\end{equation}

\noindent
here $\gamma$ is a free electron gyromagnetic ratio and $U_A$ is the energy of anisotropy. Constants $I_i\geq 0$ are related to susceptibilities as $\vec M=\frac{\partial {\cal L}}{\partial \vec H}$: magnetic susceptibilities for the field applied along i-th vector are $\chi_1=\gamma^2(I_2+I_3)$, $\chi_2=\gamma^2(I_1+I_3)$, $\chi_3=\gamma^2(I_1+I_2)$.

Anisotropy energy should be invariant under crystal symmetry transformation, its exact form depends on the symmetry of the particular crystal and on the exchange symmetry of the ordered phase, relationship between $I_i$ constants is also fixed by symmetry of the susceptibilities tensor for a given spin structure. Some examples for the known analytically solvable cases are given in Sec.\ref{sec:defs}. Note, that $I_i$ constants and exact form of the anisotropy energy are the only parameters of this approach. Once they are deduced only the formal operations remains.

Firstly, static equilibrium position $\vec{l}_i^{~(0)}$ have to be found by minimization of potential energy density

\begin{equation}
{\Pi}=-\sum_i \frac{I_i}{2}  \gamma^2 \left[\vec{ l}_i \times
  \vec{ H} \right]^2 + U_A(\left\lbrace {\vec l}_i\right \rbrace)
\end{equation}

Secondly, frequencies of small oscillations near equilibrium have to be deduced. We suppose here that these oscillations are parameterized by some three non-degenerate variables $\left \lbrace \phi_\alpha\right\rbrace$, e.g. Euler angles or other suitable variables. For the sake of simplicity we take that all of $\phi_\alpha=0$ at equilibrium position. Potential energy has a quadratic minimum at the equilibrium, thus when looking for small oscillations we can replace potential energy by its quadratic expansion. This substitution explicitly excludes possible problems of a numeric algorithm due to the finite accuracy of minimum determination.  Lagrangian density is then

\begin{equation}
{\cal L}=\sum_i \frac{I_i}{2}  \left(\dot{\vec{l}}_i\right)^2 +\gamma\sum_i I_i\left({\dot{\vec l}}_i\cdot\left[{\vec l}_i\times\vec H\right]\right) -\frac{1}{2}\sum_{\beta,\delta}\left( \frac{\partial^2 \Pi} {\partial\phi_\beta\partial\phi_\delta}\right)_{0}\phi_\beta\phi_\delta
\end{equation}
\noindent here $(...)_0$ index means that derivative is calculated at equilibrium position.

To obtain dynamics equations linear in $\phi_\alpha$ or its time derivatives, $\vec{l}_i$ have to be expanded up to second order in $\phi_\alpha$:

\begin{eqnarray*}
\vec{l}_i&=&\vec{l}_i^{~(0)}+
\sum_\beta
\left(\frac{\partial\vec{l}_i}{\partial\phi_\beta}\right)_0\phi_\beta+\frac{1}{2}\sum_{\beta,\delta}\left(\frac{\partial^2\vec{l}_i}{\partial\phi_\beta\partial\phi_\delta}\right)_0\phi_\beta\phi_\delta\\
\dot{\vec l}_i&=&\sum_\beta \left(\frac{\partial {\vec l}_i}{\partial \phi_\beta}\right)_{0} \dot{\phi}_\beta+\sum_{\beta,\delta}\left(\frac{\partial^2\vec{l}_i}{\partial\phi_\beta\partial\phi_\delta}\right)_0\phi_\beta\dot{\phi}_\delta,
\end{eqnarray*}
\noindent then with linear over $\phi_\alpha$ accuracy

\begin{eqnarray*}
\frac{\partial {\cal L}}{\partial \phi_\alpha}&=&\gamma \sum_{i, \beta\neq\alpha} \left(\left(\frac{\partial{\vec l}_i}{\partial \phi_\beta}\right)_0 \cdot \left[\left(\frac{\partial{{\vec l}_i}}{\partial \phi_\alpha}\right)_0\times\vec H\right]\right){\dot \phi}_\beta+\\
 &&+\gamma \sum_{i,\beta} I_i \dot{\phi}_\beta \left(\frac{\partial^2\vec{l}_i}{\partial\phi_\alpha\partial\phi_\beta}\right)_0\cdot\left[\vec{l}_i^{~(0)}\times\vec{H}\right]-
 \sum_\beta \left( \frac{\partial^2 \Pi} {\partial\phi_\alpha\partial\phi_\beta}\right)_{0}\phi_\beta
\end{eqnarray*}
\noindent and so forth.

Variation of the action results in three Euler-Lagrange equations

\begin{equation}
\frac{d }{dt}\frac{\partial {\cal L}}{\partial\dot{\phi}_\alpha}-\frac{\partial {\cal L}}{\partial \phi_\alpha}=0
\end{equation}

By summing up all terms and by substituting uniform harmonic oscillations $\phi_\beta=\phi_\beta^{(0)} e^{\imath\omega t}$ we obtain equations on oscillations amplitudes $\phi_\beta^{(0)}$. Required degeneracy of these equations results in the condition $det{\cal M}=0$ where matrix ${\cal M}$ of the linear equations is defined as
\begin{eqnarray}
{\cal M}_{\alpha\beta}&=&-\omega^2 \sum_i I_i \left(\left(\frac{\partial {\vec l}_i}{\partial \phi_\alpha}\right)_{0}
\cdot \left(\frac{\partial {\vec l}_i}{\partial \phi_\beta}\right)_{0}\right)+\nonumber\\
&&+2\imath\omega\gamma\sum_i I_i \left(\left(\frac{\partial{\vec l}_i}{\partial \phi_\alpha}\right)_0 \cdot \left[\left(\frac{\partial{{\vec l}_i}}{\partial \phi_\beta}\right)_0\times\vec H\right]\right)+\left( \frac{\partial^2 \Pi} {\partial\phi_\alpha\partial\phi_\beta}\right)_{0}\label{eqn:M_matrix}
\end{eqnarray}

The equation $det{\cal M}=0$ results in real cubic equation for $\omega^2$, all complex coefficients will sum to zero. Solution of this equation yields eigenfrequencies of small oscillations we sought for.

Experimental observation of these small oscillations in standard magnetic resonance experiment is, in fact, observation of the absorption of microwave radiation of certain polarization. Thus, information about oscillation of magnetization $\vec{m}(t)=\vec{m}e^{\imath\omega t}$ is important as well. It can be calculated straightforwardly as $\vec{M}=\frac{\partial{\cal L}}{\partial \vec{H}}=\vec{M}_0+\vec{m} e^{\imath \omega t}$, oscillating magnetization vector is

\begin{eqnarray}
\vec{m}&=&\imath \gamma\omega \sum_{i,\beta}I_i
\left[\left(\frac{\partial\vec{l}_i}{\partial\phi_\beta}\right)_0\times\vec{l}_i^{~(0)}\right]\phi_\beta^{(0)}-\nonumber\\
&&-
\gamma^2 \sum_{i,\beta}I_i\left(\left(\left(\frac{\partial\vec{l}_i}{\partial\phi_\beta}\right)_0\cdot \vec{H}\right)\vec{l}_i^{~(0)}+\left(\vec{l}_i^{~(0)}\cdot \vec H\right)\left(\frac{\partial\vec{l}_i}{\partial\phi_\beta}\right)_0\right)\phi_\beta^{(0)}\label{eqn:osci-m}
\end{eqnarray}
\noindent
complex form of $\vec{m}$  describes circular or elliptical precession of magnetization: $\vec{m}(t)=\left(\vec{u}+\imath\vec{v}\right)e^{\imath\omega t}$ means that real magnetization is $\vec{u} \cos\omega t-\vec {v} \sin\omega t$. Average square of longitudinal and transverse components of the oscillating magnetization can be used as a simple indicator of excitation conditions

\begin{eqnarray}
\langle {\vec m}^2\rangle&=&\frac{1}{2}\left({\vec u}^2+{\vec v}^2\right)\\
\langle m_{||}^2 \rangle&=&\frac{1}{2}\left(\left(\vec u\cdot \vec n\right)^2+\left(\vec  v\cdot \vec n\right)^2\right)\\
\langle m_{\perp}^2\rangle&=&\langle {\vec m}^2\rangle-\langle m_{||}^2 \rangle
\end{eqnarray}

\noindent here $\vec n$ is a unitary vector in the applied field direction. Being interested in the polarization of oscillating magnetization only we will norm its square averaged (if non zero) to unity: $\langle {\vec m}^2\rangle=1$.

Determination of the initial guesses for the model parameters is case-dependent. We will note here, that equation $det{\cal M}=0$ allows to scale all parameters of ${\cal M}$ arbitrary. This means, that (unless one is particulary interested to reproduce both static and dynamical properties without scaling coefficients) one of the coefficients (one of $I_i$ constants or one of the coefficients in anisotropy energy expansion) can be set to unity for convenience. Secondly, the ${\cal M}$ matrix simplifies for zero-field problem (its complex part vanishes) which could help to find zero-field gaps in AFMR spectrum. Another possible simplification is softening of the AFMR modes, which commonly appears at spin reorientation transition. In this case $\omega=0$ and $det{\cal M}=0$ reduces to $det \left( \frac{\partial^2 \Pi} {\partial\phi_\alpha\partial\phi_\beta}\right)_{0}=0$.  Finally, at high fields one of the AFMR modes is field independent and its frequency can be calculated \cite{farmar-HF}, while field-dependent modes linear asymptotes are (we assume that $\chi_3=\gamma^2(I_1+I_2)$ is the largest susceptibility)
\begin{eqnarray*}
\omega_2&=&\gamma H\\
\omega_3&=&\sqrt{1-2\frac{I_3(I_1+I_2)}{(I_1+I_3)(I_2+I_3)}}\gamma H=\sqrt{1-\frac{(\chi_1+\chi_2-\chi_3)\chi_3}{\chi_1\chi_2}}\gamma H
\end{eqnarray*}
\noindent  In the limiting case of $I_1=I_2$ ($\chi_1=\chi_2<\chi_3$) $\omega_3=\frac{I_1-I_3}{I_1+I_3}\gamma H = \frac{\chi_3-\chi_1}{\chi_1} \gamma H$. Alternatively, $I_i$ constants can be deduced from the susceptibility measurements.

\section{Solving dynamics equations numerically}
\subsection{Search for equilibrium}
We define orientation of $\left\lbrace{\vec l}_i\right\rbrace$ vectors by Euler angles $\theta$, $\phi$ and $\psi$. Minimization can be performed with any suitable standard numeric minimization procedure. However please note that numeric procedures always look for local minimum. Thus to find a global minimum one have to perform preliminary search for a starting approximation with minimal potential energy $\Pi$ over some grid in the Euler angles space. On the other hand, it could be of interest to follow a particular local minimum evolution with field, which allows to model response from different magnetic domains. MatLab implementation uses global minimum search only, C++ implementation allows to follow local minimum on user choice.

From this point on we assume that desired equilibrium position $\left\lbrace{\vec l}_i^{~(0)}\right\rbrace$ is found.
Dynamics equations are obtained by varying action $S=\int {\cal L} dV dt$ and they can be written down in any suitable variables.
Euler angles are, generally, not the best choice for dynamics equation as they suffer from ``gimbal lock'' problem: one of the degrees of freedom will be lost if at some moment ${\vec l}_3||Z$. To avoid this problem we used two approaches for calculation of eigenfrequencies: (i) to recalculate our problem to the frame of reference which is definitely free from the ``gimbal lock'', or (ii) to use other set of variables for dynamics equations. First approach was implemented in MatLab code, second approach was implemented in C++ code.

\subsection{Solving dynamics equation, MatLab implementation details}
First approach was applied in MatLab environment using the Symbolic Math Toolbox, as it provides functions for manipulating symbolic math equations and lets analytically perform differentiation, simplification and transforms. All these opportunities allow  to consider general form of the anisotropy energy $U_A$ without any simplifications. GlobalSearch class is used as well for obtaining global minimum point of potential energy $\Pi$ and finding equilibrium position $\left\lbrace{\vec l}_i^{~(0)}\right\rbrace$.

Firstly, we rotate laboratory reference frame in such a way that $\theta=\phi=\psi=\pi/6$ for equilibrium position of $\left\lbrace{\vec l}_i\right\rbrace$ vectors. The choice of angle equal to $\pi/6$ is fairly arbitrary, it is chosen simply to exclude ``gimbal lock'' problem.   Herewith recalculation of vector components of the external magnetic field and transformation of the anisotropy energy to new coordinates is needed. If $A = \left\lbrace a_{\alpha \beta} \right\rbrace$ is the matrix of this rotation, $B = A^{-1} = \left\lbrace b_{\alpha \beta} \right\rbrace$ is the inverse matrix, then in new frame of references
\begin{equation} \label{eqn:Htrans}
H^{'}_{\alpha} = \sum_\beta a_{\alpha \beta} H_{\beta}
\end{equation}
\begin{equation} \label{eqn:Utrans}
\tilde U_A(\left\lbrace l^{\alpha}_i  \right \rbrace) = U_A \Bigl( \Bigl\lbrace    \sum_\beta b_{\alpha \beta} l^{\beta}_i   \Bigr\rbrace \Bigr)
\end{equation}
Here $\vec{H} = \left\lbrace H_{\alpha} \right\rbrace$ and $\vec{H'} = \bigl\lbrace H^{'}_{\alpha} \bigr\rbrace$ are vectors of the external magnetic field in the basic and transformed frames of references correspondingly, $\tilde U_A(\left\lbrace l^{\alpha}_i  \right \rbrace)$ is the anisotropy energy written in new frame of references.

Secondly, we use parametrization of Euler angles for description of small oscillations near the equilibrium position in transformed frame of references, because in such case ``gimbal lock'' problem is avoided. As magnetic vectors components $\left\lbrace l^{\alpha}_i  \right \rbrace$ are known functions of $\theta$, $\phi$, $\psi$ parameters, there are no any problems to obtain the values of first derivatives of $\left\lbrace l^{\alpha}_i  \right \rbrace$ vectors and the values of first and second derivatives of potential energy $\Pi$ at $\left\lbrace{\vec l}_i^{~(0)}\right\rbrace$ position. These values are used for calculations of oscillations eigenfrequencies from the equation $det{\cal M}=0$ according to Eqn.(\ref{eqn:M_matrix}).

Complete algorithm is divided into few steps:
\begin{enumerate}
\item We start from specified start field $H=H_{start}$ applied in the specified direction.
\item We look for global minimum of potential energy $\Pi$ and find a new equilibrium position at field $H$. Information on equilibrium position (Euler angles, potential energy at equilibrium, projections of $\left\lbrace\vec{l}_i^{~(0)}\right\rbrace$ vectors on the field direction, longitudinal and transverse susceptibilities) is saved.
\item Components of vector $\vec{H'}$ (Eqn.\ref{eqn:Htrans}) and anisotropy energy $\tilde U_A$ (Eqn.\ref{eqn:Utrans}) in transformed frame of references are obtained.
\item Matrix ${\cal M}$ (Eqn.\ref{eqn:M_matrix}) is calculated and $det{\cal M}=0$ equation is solved for eigenfrequencies. Results are saved.
\item Eigenvectors and average values of projections of oscillating magnetization vector along and transverse to external magnetic field for all oscillation modes are found and saved.
\item Field is increased by specified increment $H_{step}$. If the field does not reach its goal value $H_{stop}$ we continue with Step 2.
\end{enumerate}

All input parameters including anisotropy energy function $U_A$ in general case, $\chi_i$ and $\gamma$ coefficients, magnetic field direction, variation boundaries, increment of the value of magnetic field are specified in MatLab script, available at \cite{home}. Calculation results are saved in three files correspondingly with static properties (equilibrium position, energy at equilibrium, projections of $\left\lbrace\vec{l}_i^{~(0)}\right\rbrace$ vectors on the field direction, longitudinal and transverse susceptibilities), oscillation eigenfrequencies and eigenvectors together with average projections of oscillating magnetization vector along and transverse to the magnetic field. Format of these files is described in details in supplementary materials (see below).

\subsection{Solving dynamics equation, C++ implementation details}
MatLab environment allows big flexibility and allow to avoid lot of routine operations. However, it requires commercial software and, being an interpreter, is somewhat slower then a properly compiled program. Thus we propose an alternative implementation in C++ language along with flexible executable program. C++ codes and compiled Win32 executable are available at \cite{home}.

For the sake of flexibility we will consider only quadratic terms in anisotropy energy

\begin{equation}\label{eqn:U-C}
 U_A=\sumprimed{i, j, \alpha,\beta} a_{ij}^{\alpha,\beta}~ l_i^{\alpha} l_j^{\beta}
\end{equation}

\noindent here $\sumprimed{}$ sign means that each $l_i^{\alpha} l_j^{\beta}$ combination is counted only once during summation. Higher orders of anisotropy can be included in the program code in a straightforward way, if necessary. This restriction allows to read all $a_{ij}^{\alpha\beta}$ coefficients from easily editable plain text ini-file (ini-file format is described in supplementary materials below) and to simplify all derivatives calculations for minimum search routine and for dynamics equation derivation. E.g.,
\begin{equation}
\frac{\partial U_A}{\partial x}=\sumprimed{i, j, \alpha,\beta} a_{ij}^{\alpha,\beta} \left(\frac{\partial  l_i^{\alpha}}{\partial x} l_j^{\beta}+ l_i^{\alpha} \frac{\partial  l_j^{\beta}}{\partial x}\right)
\end{equation}
\noindent here $x$ is some variable of choice.

Numerical Recipes \cite{NR} \textsc{frprmn} routine is used to find an equilibrium position. We continue calculations in the same frame of references attached to the crystal, but small oscillations near the equilibrium are described as a small rotations of  $\left\lbrace{\vec l}_i\right\rbrace$ vectors parameterized by a vector of small rotations $\vec \phi=(\phi_x,\phi_y,\phi_z)$. Length of this vector is rotation angle and its direction defines rotation axis, at equilibrium position $\phi=0$. Up to quadratic terms in $\phi$ transformation of $\left\lbrace{\vec l}_i\right\rbrace$ can be described as:
\begin{equation}
   {\vec l}_i={\vec l}_i^{(0)}+\left[\vec{\phi}\times\vec{l}_i^{~(0)}\right]+\frac{1}{2}\left[\vec{\phi}\times\left[\vec{\phi}\times\vec{l}_i^{~(0)}\right]\right]+\underline{O}\left(\phi^3\right)
\end{equation}
\noindent
This parametrization is free from ``gimbal lock''. Note that there are nonzero second order derivatives $\frac{\partial^2\vec{l}_i}{\partial\phi_\alpha\partial\phi_\beta}$ which have to be taken into account when  calculating Hessian matrix $\frac{\partial^2 \Pi} {\partial\phi_\alpha\partial\phi_\beta}$. This allows to complete calculations of oscillations eigenfrequencies.

Once eigenfrequencies are known, complex oscillation vectors $\vec{\phi}$ are found as zero-eigenvalue eigenvectors of ${\cal M}$ matrix using standard \textsc{jacobi} procedure from Numerical Recipes \cite{NR}. This allows to compute complex oscillating magnetization vector  $\vec m$ (see Eqn.(\ref{eqn:osci-m})) and its average projections on the field direction and on the direction transverse to the field.

Complete algorithm looks as follows:
\begin{enumerate}
\item We start from specified start field $H=H_{start}$ applied in the specified direction.
\item We look for a new equilibrium position at field $H$. According to user choice we either look for global minimum or for a local minimum close to some initial approximation (specified initial approximation at first point or previous equilibrium position). Information on equilibrium position (Euler angles, projections of ${\vec l}_i$ on the field direction and longitudinal susceptibility) is saved.
\item Matrix ${\cal M}$ (Eqn.\ref{eqn:M_matrix}) is calculated and $det{\cal M}=0$ equation is solved for eigenfrequencies. Results are saved.
\item Oscillating complex magnetization components and average longitudinal and transverse components of the oscillating magnetization for all oscillation modes are found and saved.
\item Field is increased by specified increment $H_{step}$. If the field does not reach its goal value $H_{stop}$ we continue with Step 2.
\end{enumerate}

All input parameters including anisotropy energy coefficients (Eqn.\ref{eqn:U-C}), $I_i$ and $\gamma$ coefficients, magnetic field direction and limiting boundaries are specified in a text ini-file. Calculation results are saved in three files with static properties (equilibrium position, energy at equilibrium, longitudinal and transverse susceptibilities), oscillation eigenfrequencies  and eigenvectors correspondingly. Formats of the ini-file and of the output files are described in details in supplementary materials below.

\subsection{Application to the test examples}

\begin{figure}
 \centering
 \epsfig{file=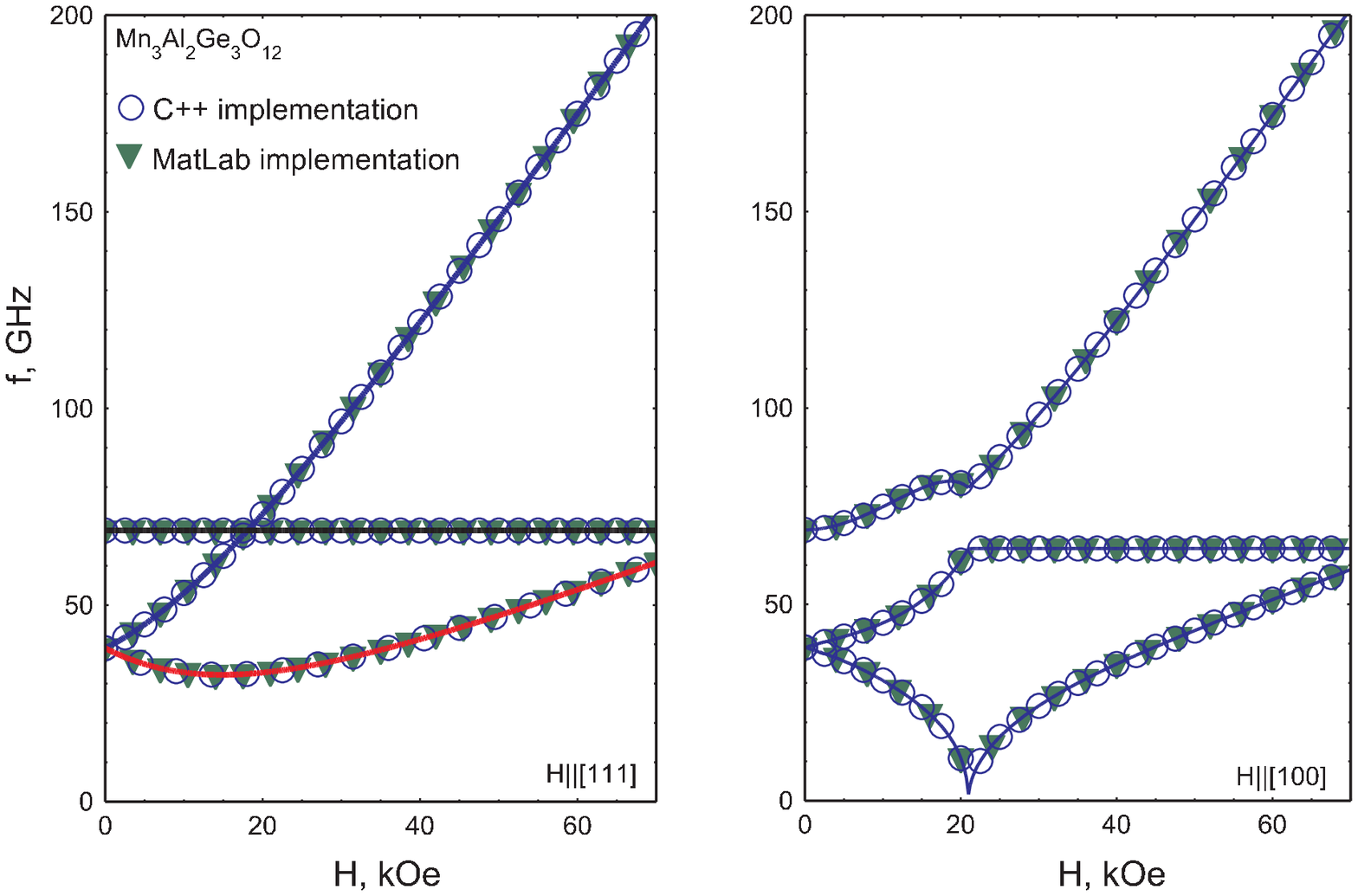, width=\figwidth, clip=}
 \caption{Application of the numeric algorithms to the test example of \AlMn{}. Model parameters reproduce 1.2K experimental data of Ref.\cite{garnet-afmr} and are listed in the Appendix. On all panels closed symbols are the results of MatLab implemented algorithm, open symbols are the results of C++ implemented algorithm. (Left panel) $\vec{H}||[111]$, bold solid lines are analytical solution; (Right panel) $\vec{H}||[100]$, curves are guide to the eye.}\label{fig:sample}
\end{figure}

We tested our algorithms against test cases described in Sec.\ref{sec:defs}. Example of the numerically computed AFMR $f(H)$ dependence is shown at the Figure \ref{fig:sample}, detailed tests protocols are included in supplementary material below.

Test routine included: application to the test cases with known analytical results for $f(H)$, computation at the equivalent field orientations for cubic crystal, computation of the $f(H)$ curve at canted field orientation. We have found that numeric results coincides with known analytical solutions, both implementations of the algorithm yield the same results, no ``gimbal lock'' cases occurs.

Some minor instabilities of the numeric procedures were noted in highly degenerate cases (coincidence of resonance frequencies for different modes or presence of a zero-frequency mode), but they affect only less important output data. We found that sometimes determination of the frequency for $\omega=0$ mode, which is not experimentally observable, is faulty or excitation condition determination is sometimes uncertain for the degenerate modes. Determination of the static properties and $f(H)$ curves for $f\neq 0$ was not affected by these issues.

\section{Conclusions (main matter)}
We present the algorithm for numerical solution of antiferromagnetic resonance frequencies for a noncollinear antiferromagnet of a general type within framework of the exchange symmetry theory \cite{andrmar}. Algorithm is implemented in the available MatLab and C++ codes (including ready-to-use compiled win32 executable) \cite{home}, implementations are tested against known analytically solvable models.

Authors thank Prof.A.I.Smirnov and Dr. L.E.Svistov (Kapitza Institute) for useful discussions. Work was supported by Russian Foundation for Basic Research grant No.16-02-00688.

\part{Supplementary materials: detailed code description, output file formats and detailed tests}

\section{Source files location}
Source files of the MatLab and C++ implementations of the numeric algorithm for description of the AFMR modes of noncollinear antiferromagnets are available through the Authors web-site \textsf{www.kapitza.ras.ru/rgroups/esrgroup/} (``NuMA: Numeric Methods for Antiferromagnets'' section) \cite{home}.

\section{MatLab implementation}
\subsection{Editable fragments of the source file}

MatLab source file \textsf{Noncolaf.m} starts from declarations of variables used and their initialization. We list here location of the editable lines in this file:

\begin{description}
\item[lines 1-2] These lines determine MatLab version used and creation date of the \textsf{Noncolaf.m} file. Following description is valid for the file dated as 21.04.2016.

\item[lines 44-47] Set field direction (azimuthal and polar angles \textsf{hphi} and \textsf{htheta} in the crystallographic frame). Set field scan parameters: low field limit \textsf{Hlow}, high field limit \textsf{Hhigh}, field increment \textsf{delta}.

\item[lines 49-62] Set parameters and form of anisotropy energy for the test case examples. Parameters and anisotropy energy for other cases should be specified in the same way. Please comment all unused parameters sets with a \% symbol in the beginning of the line.

\end{description}

Selection of units is the user choice. In the case of the test examples magnetic field units are kOe, gyromagnetic ratio $\gamma$ units are $\frac{10^9 rad \cdot s^{-1}}{kOe}$ (which allows to obtain final frequencies as GHz), $I_i$ units are $\frac{ kOe^2}{(10^9 rad\cdot s^{-1})^2}$. We set one of the coefficients of anisotropy energy to unity, units of potential energy density are then $kOe^2$.

\subsection{Output files format}
Oscillations eigenfrequencies are saved in the default file ``Oscillation Eigenfrequencies.txt''. This is 4-column txt-file:
\begin{description}
\item[col.1] Magnetic field value;
\item[col.2-4]  Oscillation frequencies.
\end{description}
\noindent
In the default set of units field is measured in kOe and frequencies are measured in GHz.

Static properties are saved in the default file ``Static Properties.txt''. This is a 10-columns txt-file:
\begin{description}
\item[col.1] Magnetic field value;
\item[col.2-4]  Euler angles $\alpha$, $\beta$ and $\gamma$ (radians) describing equilibrium position of $\left\lbrace\vec{l}_i^{~(0)}\right\rbrace$ vectors;
\item[col.5] Potential energy value at equilibrium;
\item[col.6-8] Projections of $\left\lbrace{\vec l}_i^{~(0)}\right\rbrace$ vectors on the field direction;
\item[col.9-10] Longitudinal and transverse magnetic susceptibilities at equilibrium ($\chi_\parallel=M_\parallel/H$, $\chi_\perp=M_\perp/H$).
\end{description}
\noindent

Information on oscillations eigenvectors and average values of longitudinal and transverse oscillating magnetization is saved in the default file ``Eigenvectors and Oscillating Magnetization Projections.txt''. This is 25-columns txt-file:

\begin{description}
\item[col.1] Magnetic field value;
\item[col.2-7] Components of the complex eigenvector for the first oscillation mode. These components are calculated in the Euler angles space in the rotated frame of reference. The order of component is as follows: $Re(\alpha)$, $Im(\alpha)$, $Re(\beta)$, $Im(\beta)$, $Re(\gamma)$, $Im(\gamma)$.
\item[col.8-19] Components of complex eigenvectors for second and third oscillation modes.
\item[col.20-25] Average squared of longitudinal and transverse oscillating magnetization ($\sqrt{\langle {\vec {m}}_\parallel^2 \rangle}$  and $\sqrt{\langle {\vec {m}}_\perp^2 \rangle}$) for all three oscillation modes. The order is as follows: $\sqrt{\langle {\vec {m}}_{1\parallel}^2 \rangle}$, $\sqrt{\langle {\vec {m}}_{1\perp}^2 \rangle}$, $\sqrt{\langle {\vec {m}}_{2\parallel}^2 \rangle}$, $\sqrt{\langle {\vec {m}}_{2\perp}^2 \rangle}$, $\sqrt{\langle {\vec {m}}_{3\parallel}^2 \rangle}$, $\sqrt{\langle {\vec {m}}_{3\perp}^2 \rangle}$.
\end{description}

\section{C++ implementation}
\subsection{Source files}
This description corresponds to the v.1.00 dated as June 28,2016 of the program code. One can get program version information by running program with \textsf{-v} command line key (e.g., for the case of compiled Win32 application provided in the package, type \textsf{noncolaf-win32.exe -v } in a command line).

Source files includes some files from Numerical Recipes package \cite{NR}: \textsf{brent.c}, \textsf{f1dim.c}, \textsf{frprmn.c}, \textsf{jacobi.c}, \textsf{linmin.c}, \textsf{mnbrac.c}, \textsf{nrutil.c}, \textsf{nr.h}, \textsf{nrutil.h}. These files contains minimization routine \textsf{frprmn.c} and eigenvector search routine \textsf{jacobi.c}. These routines perform calculation with standard float precision, which is found to suffice for our goals.

Main program file is \textsf{noncolaf.cpp}, it uses additional functions defined in the set of header files:
\begin{description}
\item[\textsf{cubic.h}] Solution of the real cubic equation
\item[\textsf{vector.h}] Defines 3D-vector algebra and reload standard operators to simplify vector operations
\item[\textsf{ini.h}] Reading INI-file
\item[\textsf{magvect.h}] Defines set of $\left\lbrace{\vec l}_i\right\rbrace$ vectors and their derivatives
\item[\textsf{energy.h}] Calculates potential energy and its derivatives
\item[\textsf{minsearch.h}] Defines functions to be called by minimization routine and preliminary search of the local minimum over the grid in the Euler angles space
\item[\textsf{saving.h}] Includes calculations of the output parameters and their saving to the appropriate files
\end{description}

Full list of the source files includes 17 files. Source files were compiled using a DevC++ compiler v5.11 into a Win32 console application \textsf{noncolaf-win32.exe} which is also available for download.

Default output file names are specified in the lines 34-37 of the main \textsf{noncolaf.cpp} file, program version is defined in line 8 of the \textsf{noncolaf.cpp} file, default INI-file name is specified in the line 94 of \textsf{ini.h} file.

Presently C++ implementation consider only quadratic invariants in the anisotropy energy

\begin{equation*}
 U_A=\sumprimed{i, j, \alpha,\beta} a_{ij}^{\alpha,\beta}~ l_i^{\alpha} l_j^{\beta}
\end{equation*}

\noindent here $\sumprimed{}$ sign means that each $l_i^{\alpha} l_j^{\beta}$ combination is counted only once during summation. If one intend to include higher order terms corresponding modifications have to be done in \textsf{energy.h} file (potential energy and its derivatives calculations).

\subsection{INI-file format}

All parameters used for calculations are read from plain text INI-file. Default INI-file name is \textsf{noncolaf.ini}, examples of this file for the test cases are included in the package. INI-file format includes headers in square brackets (e.g., \textsf{[gamma:]}) followed by the line with numeric value of appropriate parameter. Order of the parameters specification in the INI file is arbitrary, however numerical value of the parameter declared by its header have to be provided prior to other header declaration.

Dummy INI-file template with all necessary headers present will be created if no ini-file will be found. Examples of the INI-files for the test examples described in Sec.\ref{sec:defs} are available for download \cite{home}. Consistency control of INI-file parameters for common errors (e.g., negative $I_i$) is performed before starting modeling, program terminates with appropriate warning if such an error is found.

List of the required INI-file headers is given below.

\noindent
\emph{Lagrangian parameters}:
    \begin{description}
        \item[\textsf{[gamma:]}] Gyromagnetic ratio $\gamma$. Default units are $\frac{10^9 rad \cdot s^{-1}}{kOe}$ (which allows to obtain final frequencies as GHz).
        \item[\textsf{[I$_1$:]}, \textsf{[I$_2$:]}, \textsf{[I$_3$:]}] Set of $I_i$ parameters. Default units are $\frac{ kOe^2}{(10^9 rad\cdot s^{-1})^2}$.
        \item[\textsf{[AnisotropyStart:]}] This header marks start of the anisotropy energy definition. It have to be followed by \textsf{[AnisotropyEnd:]} later (it is the only header that have to be terminated by other header). Anisotropy energy is assumed in the form $U_A=\sumprimed{i, j, \alpha,\beta} a_{ij}^{\alpha,\beta}~ l_i^{\alpha} l_j^{\beta}$, all $a_{ij}^{\alpha,\beta}~ l_i^{\alpha} l_j^{\beta}$ terms have to be included between \textsf{[AnisotropyStart:]} and \textsf{[AnisotropyEnd:]} as separate lines. Format is straightforward and can be illustrated by the example of \AlMn{}: in this case $U_A=\lambda \left[l_{2z}^2-l_{1z}^2+\frac{2}{\sqrt{3}}\left(l_{1x}l_{2x}-l_{1y}l_{2y}\right)\right]$ and corresponding INI-file fragment looks as

            \begin{tabular}{l}
            \textsf{[AnisotropyStart:]}\\
            \textsf{l2zl2z;1}\\
            \textsf{l1zl1z;-1}\\
            \textsf{l1xl2x;1.15470054} \\
            \textsf{l1yl2y;-1.15470054} \\
            \textsf{[AnisotropyEnd:]}
            \end{tabular}

            \noindent
            I.e. each line is a description of $l_i^{\alpha} l_j^{\beta}$ term followed by numerical value of coefficient, separated by semicolon. Each $l_i^{\alpha} l_j^{\beta}$ combination should appear only once, otherwise error message appears and program terminates.
    \end{description}

\noindent
\emph{Field scan parameters}:
    \begin{description}

    \item[\textsf{[Hdir:]}] Specifies field direction as semicolon separated 3D-vector, e.g.:

        \begin{tabular}{l}
        \textsf{[Hdir:] (Semicolon separated vector)}\\
        \textsf{0;0;1}
        \end{tabular}

    \noindent The length of this vector is arbitrary, program will norm it to unity during operation.
    \item[\textsf{[Hstart:]}, \textsf{[Hstop:]}, \textsf{[Hstep:]}] Starting and final field values and field increment. Field scan can be modeled both on the increasing or on the decreasing field (\textsf{[Hstep:]} should be negative in the later case). Default field units are kOe.
    \end{description}

\noindent
\emph{Algorithm versatility parameters}:
    \begin{description}
    \item[\textsf{[minsearch flag:]}] Should be 1 or 0. If set to 1 then global minimum search over Euler angles space is performed when looking for $\left\lbrace{\vec l}_i\right\rbrace$ equilibrium orientation If set to 0 then local minimum is followed: specified initial approximation is used on the first step, equilibrium position found is used as an initial approximation on the next step and so on.

    \item[\textsf{[grid size:]}] Should be integer $N$. It determines grid size in the Euler angles space that is used for rough equilibrium search if ``global search'' option is selected (\textsf{[minsearch flag:]} header followed by 1). Rough search includes estimation of the potential energy in the $N^3$ points regularly spaced in Euler angles space. Recommended value is from 10 to 30.

    \item[\textsf{[starting approximation:]}] Determines Euler angles of the approximate equilibrium position used at the first point if ``local search'' option is selected (\textsf{[minsearch flag:]} header followed by 0). Euler angles are defined as semicolon separated line: $\Theta;\phi;\psi$.
    \end{description}

\subsection{Error handling}

Program checks INI-file for consistency and checks validity of some parameters (positiveness of $\gamma$ and $I_i$, consistency of field scan parameters, unique definitions of anisotropy energy terms).

During calculations numeric uncertainties can result in incorrect results especially in strongly degenerated cases (usually if zero oscillation frequency is present). Sometimes this results in small negative or even complex $\omega^2$ roots of $det{\cal M}=0$ equation. We arbitrary set a small cutoff limit (line 11 of \textsf{saving.h}) with default value $-1\cdot 10^{-4}$, negative $\omega^2$ above $(2\pi)^2$ times cutoff value is set to zero allowing for numeric uncertainty. In other cases (complex roots or larger negative $\omega^2$ corresponding output frequency is set to $-1$ and oscillation eigenvectors are set to zero.

\subsection{Output files format}

Program creates 4 output files, default names are \textsf{noncolaf.dsk}, \textsf{noncolaf.st}, \textsf{noncolaf.frq} and \textsf{noncolaf.mag}. All of output data files includes header describing briefly its contents.

File \textsf{noncolaf.dsk} contains information about modeling parameters. It essentially duplicates INI file, presenting the same information in a more friendly formatted way.

File \textsf{noncolaf.st} contains information about static properties at equilibrium position. It is a 10 column text file:
    \begin{description}
        \item[col.1] Magnetic field
        \item[col.2] Potential energy value at equilibrium;
        \item[col.3-4] Longitudinal and transverse magnetic susceptibilities at equilibrium ($\chi_\parallel=M_\parallel/H$, $\chi_\perp=M_\perp/H$).
        \item[col.5-7]  Euler angles $\Theta$, $\phi$ and $\psi$ (radians) describing equilibrium position of $\left\lbrace\vec{l}_i^{~(0)}\right\rbrace$ vectors;
        \item[col.8-10] Directing cosines of magnetic field with respect to $\left\lbrace{\vec l}_i^{~(0)}\right\rbrace$ vectors (i.e., $({\vec l}_i^{~(0)} \cdot \vec {n})$, here $\vec n$ is a unitary vector in the field direction).
    \end{description}

File \textsf{noncolaf.frq} contains oscillation frequencies. It is a 4 column txt-file:
    \begin{description}
        \item[col.1] Magnetic field value;
        \item[col.2-4]  Oscillation frequencies.
    \end{description}

File \textsf{noncolaf.mag} contains information on oscillation eigenvectors and average values of longitudinal and transverse components of oscillating magnetization. It is a 10-column txt-file:
    \begin{description}
        \item[col.1] Magnetic field value;
        \item[col.2-7]  Average squared of longitudinal and transverse oscillating magnetization ($\sqrt{\langle {\vec {m}}_\parallel^2 \rangle}$  and $\sqrt{\langle {\vec {m}}_\perp^2 \rangle}$) for all three oscillation modes. The order is as follows: $\sqrt{\langle {\vec {m}}_{1\perp}^2 \rangle}$, $\sqrt{\langle {\vec {m}}_{1\parallel}^2 \rangle}$, $\sqrt{\langle {\vec {m}}_{2\perp}^2 \rangle}$, $\sqrt{\langle {\vec {m}}_{2\parallel}^2 \rangle}$, $\sqrt{\langle {\vec {m}}_{3\perp}^2 \rangle}$ $\sqrt{\langle {\vec {m}}_{3\parallel}^2 \rangle}$;
        \item[col.8-10] Oscillation eigenvectors for all modes. Each vector is a real space complex vector, each vector components are semicolon separated as follows: $[Re(m_x)+\imath \cdot Im(m_x)];[Re(m_y)+\imath \cdot Im (m_y)];[Re (m_z)+\imath \cdot Im (m_z)]$
    \end{description}

\section{Detailed test protocols}
\subsection{Analytically solvable models used as a test cases\label{sec:defs}}

We recall here some of the known examples of application of exchange symmetry theory to low-energy dynamics of noncollinear antiferromagnets. These analytical solutions were used as a test cases to ascertain correctness of numeric algorithms.

First test example is an antiferromagnet on a triangular lattice \csnicl{} \cite{csnicl-afmr}. In the ordered phase of this magnet spins form a planar 120$^\circ$ structure.  High symmetry of triangular lattice leaves single invariant in the anisotropy energy $U_A=\beta \left({l_{3}^z}\right)^2$, here $z$ axis is normal to hexagonal plane and vector ${\vec l}_3$ is the normal to the plane of the planar spin structure, $\beta>0$ as at zero field spin plane is orthogonal to the hexagonal crystallographic plane. Magnetic susceptibility normal to the spin plane dominates: $\chi_3>\chi_2=\chi_1$ (i.e. $I_3<I_1=I_2$). Two of the zero-field frequencies are zero, nonzero zero-field frequency is $\omega_{0}=\gamma \sqrt{\frac{I_1-I_3}{I_1+I_3}\beta}=\gamma\sqrt{\frac{\chi_3-\chi_1}{\chi_1}\beta}$. As the field is applied along $z$ axis spin plane reorients at the field $H_0=\sqrt{\frac{\beta}{\gamma^2 (I_1-I_3)}}=\sqrt{\frac{\beta}{\chi_3-\chi_1}}$. Magnetic resonance frequencies at $\vec{H}||z$ are given by equations:

\begin{eqnarray*}
H<H_0&:& \omega_1^2=\omega_{0}^2+\left(\gamma H\right)^2\\
    & & \omega_2=\omega_3=0\\
H>H_0&:&\omega_{1,2}=\sqrt{\left(\frac{I_1}{I_1+I_3} \gamma H \right)^2-\omega_{10}^2}\pm \frac{I_3}{I_1+I_3}\gamma H=\\
 & &=\sqrt{\left(\frac{\chi_3}{2\chi_1}\gamma H \right)^2-\omega_{0}^2}\pm \frac{2\chi_1-\chi_3}{2\chi_1}\gamma H\\
 & &\omega_3=0
\end{eqnarray*}

\noindent Because of simplicity of anisotropy energy this problem can be solved analytically at arbitrary field orientation, see Ref.\cite{csnicl-afmr} for details.

To reproduce experimental results of Ref.\cite{csnicl-afmr} we take for our modeling  $\beta=1$ kOe$^2$, $\gamma=18.8  \frac{10^9 rad\cdot s^{-1}}{kOe}$ ($3.0$ GHz/kOe in frequency units), $I_1=I_2=8.77\cdot10^{-6} \frac{ kOe^2}{(10^9 rad\cdot s^{-1})^2}$ and $I_3=9.75\cdot10^{-7}\frac{ kOe^2}{(10^9 rad\cdot s^{-1})^2}$.

Secondly, we consider twelve-sublattices antiferromagnet \AlMn{} \cite{garnet-afmr}. Here $I_1=I_2$ because of the cubic symmetry, anisotropy energy $U_A=\lambda \left[l_{2z}^2-l_{1z}^2+\frac{2}{\sqrt{3}}\left(l_{1x}l_{2x}-l_{1y}l_{2y}\right)\right]$ ($\lambda>0$) (we use notations of Ref.\cite{udalov}). At zero field plane of the spiral structure is orthogonal to one of the $\langle 111\rangle$ directions. Oscillation eigenfrequencies can be found at $\vec{H}||[111]$:

\begin{eqnarray*}
\omega_{1,2}&=&\sqrt{ \left(\frac{I_1}{I_1+I_3}\gamma H\right)^2+\frac{4}{3}\frac{\lambda}{(I_1+I_3)}}
\pm \frac{I_3}{I_1+I_3}\gamma H=\\
 &=&\sqrt{ \left(\frac{\chi_3}{2\chi_1}\gamma H\right)^2+\frac{4}{3}\frac{\lambda}{\chi_1}\gamma^2}
\pm \frac{2\chi_1-\chi_3}{2\chi_1}\gamma H \\
\omega_3&=&\sqrt{\frac{8}{3}\frac{\lambda}{I_1}}=\gamma \sqrt{\frac{8}{3}\frac{2\lambda}{\chi_3}}
\end{eqnarray*}

To reproduce experimental results  of Ref. \cite{garnet-afmr} we take for our modeling $\lambda=1$ kOe$^2$, $\gamma=17.6\frac{10^9 rad\cdot s^{-1}}{kOe}$ ($2.80$ GHz/kOe), $I_1=I_2=1.42\cdot 10^{-5}\frac{ kOe^2}{(10^9 rad\cdot s^{-1})^2}$, $I_3=7.99\cdot 10^{-6}\frac{ kOe^2}{(10^9 rad\cdot s^{-1})^2}$. Results of the modeling for this case are shown at the Figure \ref{fig:sample}.

Finally, it is a spiral magnet \licuo{} \cite{svistov-licuo}. Despite of the orthorhombic symmetry $I_1=I_2$ as there is no anisotropy in the plane of the spiral structure, $U_A=\frac{A}{2}l_{3z}^2+\frac {B}{2} l_{3y}^2$ ($A\leq B \leq 0$). It turns out that in the case of \licuo{} $A$ and $B$ constants in anisotropy energy are  close within 1\%. Thus, normal to the spin plane $\vec{l}_3$ rotates almost freely in the $(yz)$ plane.   One of the oscillation frequencies corresponds to the rotation in the plane of spiral structure and is always zero since phase of the helix can be changed at no energy cost. Two other modes have non-zero zero-field frequencies

\begin{eqnarray*}
\omega_{10}^2&=&-\frac{A}{I_1+I_3}=-\gamma^2 \frac{A}{\chi_1}\\
\omega_{20}^2&=&\frac{B-A}{I_1+I_3}=\gamma^2\frac{B-A}{\chi_1}<\omega_{10}^2\\
\end{eqnarray*}

For \licuo{} $\chi_3>\chi_1$, in this case at $\vec{H}||z$ vector $\vec{l}_3$ always remains aligned along $z$ and non-zero oscillation frequencies are
\begin{eqnarray*}
\omega_{1,2}^2&=&\frac{\omega_{10}^2+\omega_{20}^2}{2}+
\gamma^2 H^2\frac{I_3^2+I_1^2}{\left(I_3+I_1\right)^2}\pm\\
&&\pm \sqrt{\left(\frac{\omega_{10}^2-\omega_{20}^2}{2}\right)^2+4\frac{\gamma^4H^4 I_1^2I_3^2}{\left(I_1+I_3\right)^4}+2\frac{\gamma^2 H^2 \left(\omega_{10}^2+
\omega_{20}^2\right)I_3^2}{\left(I_1+I_3\right)^2}}
\end{eqnarray*}

At $\vec{H}||x$ spin plane rotates orthogonally to the magnetic field at some critical field. Critical field $H_{cx}=\frac{\omega_{10}}{\gamma}\sqrt{\frac{I_1+I_3}{I_1-I_3}}=\frac{\omega_{10}}{\gamma}\sqrt{\frac{\chi_1}{\chi_3-\chi_1}}$ and oscillation frequencies are

\begin{eqnarray*}
H<H_{cx}&:&\\
\omega_1^2&=&\omega_{10}^2+\gamma^2 H^2\\
\omega_2^2&=&\omega_{20}^2\\
H>H_{cx}&:&\\
\omega_{1,2}^2&=&\frac{\omega_{20}^2-2\omega_{10}^2}{2}
+\frac{I_1^2+I_3^2}{\left(I_1+I_3\right)^2}\gamma^2 H^2 \pm\\
&&\pm\sqrt{\frac{\omega_{20}^4}{4}+
2\gamma^2 H^2\frac{\left(\omega_{20}^2-2\omega_{10}^2\right)I_3^2}
{\left(I_1+I_3\right)^2}
+4\frac{\gamma^4 H^4 I_1^2 I_3^2}{\left(I_1+I_3\right)^4}}
\end{eqnarray*}

To reproduce experimental results of Ref. \cite{svistov-licuo} we take for our modeling $\gamma=17.59\frac{10^9 rad\cdot s^{-1}}{kOe}$ (corresponds to 2.80 GHz/kOe), $A=-1$kOe$^2$, $B=-0.99$ kOe$^2$, $I_1=I_2=1.85\cdot 10^{-7}\frac{ kOe^2}{(10^9 rad\cdot s^{-1})^2}$, $I_3=6.18\cdot 10^{-8}\frac{ kOe^2}{(10^9 rad\cdot s^{-1})^2}$

\subsection{Case of \csnicl{}}
\begin{figure}
  \centering
 \epsfig{file=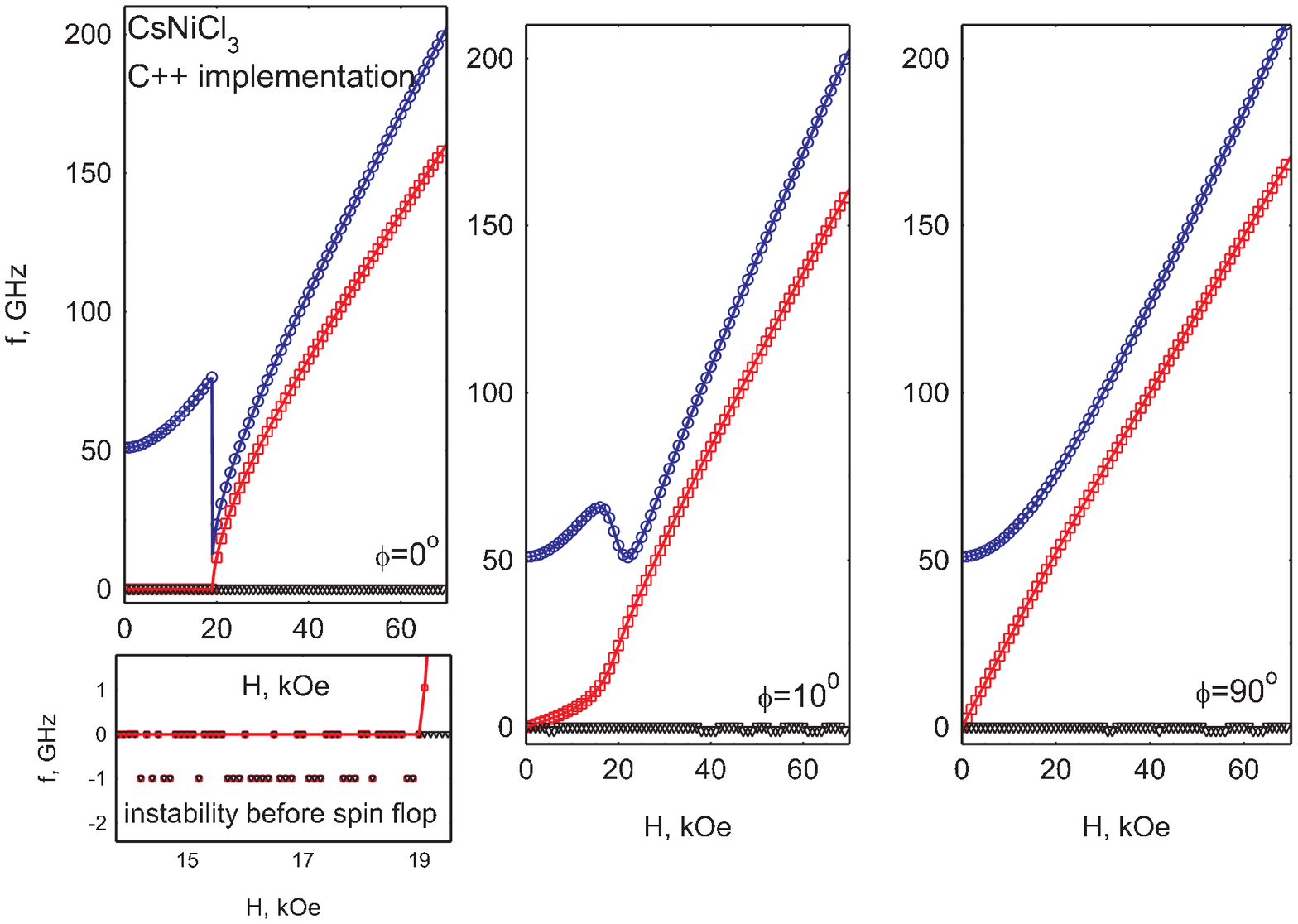, width=\figwidth, clip=}
  \caption{Frequency-field dependences modeled for\csnicl{} using C++ implementation of the numeric procedure. Curves are analytical results, symbols --- numeric modeling. Main panels show $f(H)$ curves at different field orientation with respect to the anisotropy axis $z$ ($\phi=0^\circ$ means $\vec{H}||z$). Small panel illustrates instability of the numeric procedure due to the strong degeneration of the dynamics equations in this case.}\label{fig:csnicl-cpp}
\end{figure}

\begin{figure}
  \centering
 \epsfig{file=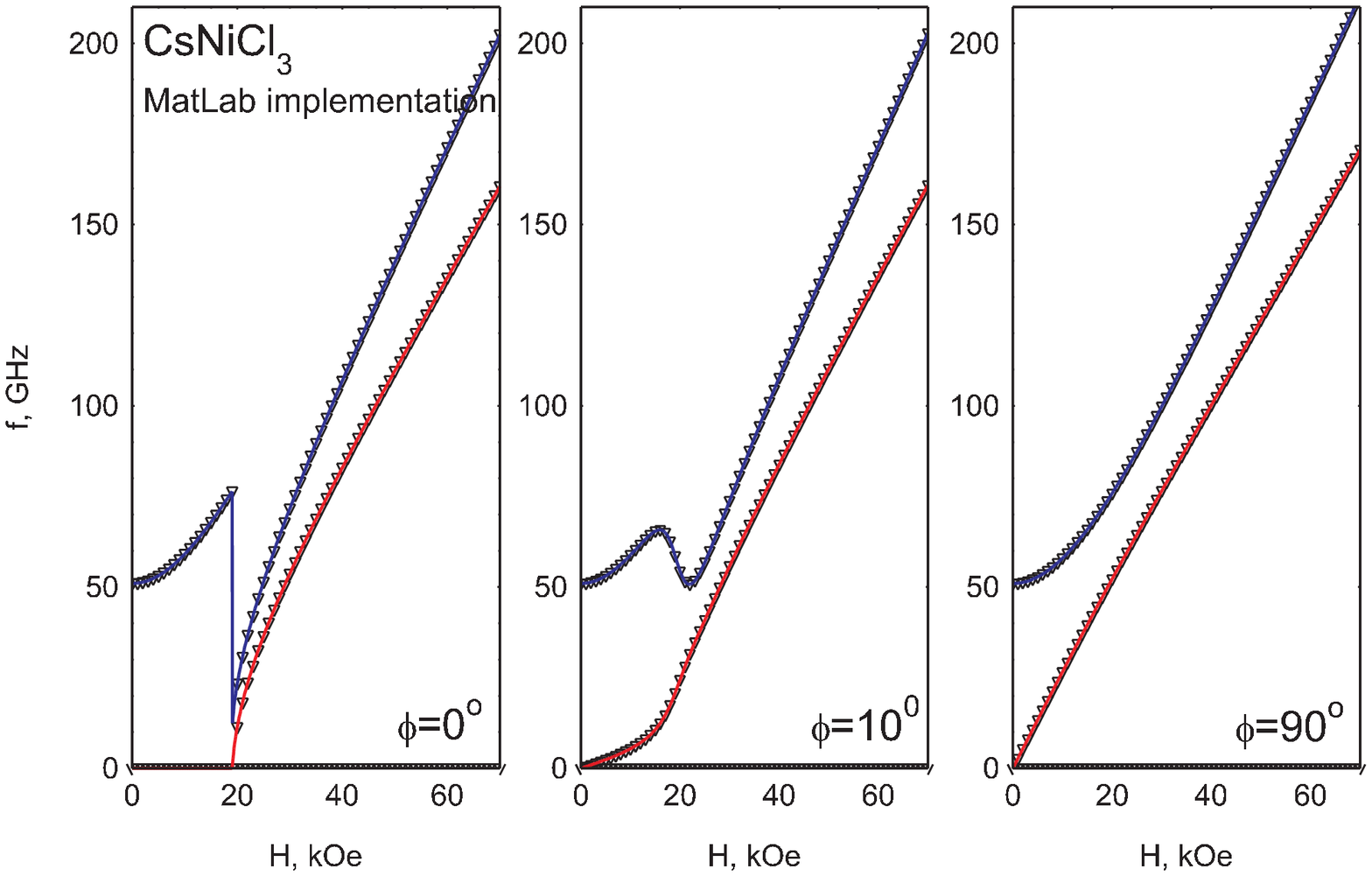, width=\figwidth, clip=}
  \caption{Frequency-field dependences modeled for\csnicl{} using MatLab implementation of  the numeric procedure. Curves are analytical results, symbols --- numeric modeling. Panels show $f(H)$ curves at different field orientation with respect to the anisotropy axis $z$ ($\phi=0^\circ$ means $\vec{H}||z$). }\label{fig:csnicl-matlab}
\end{figure}

\begin{figure}
  \centering
 \epsfig{file=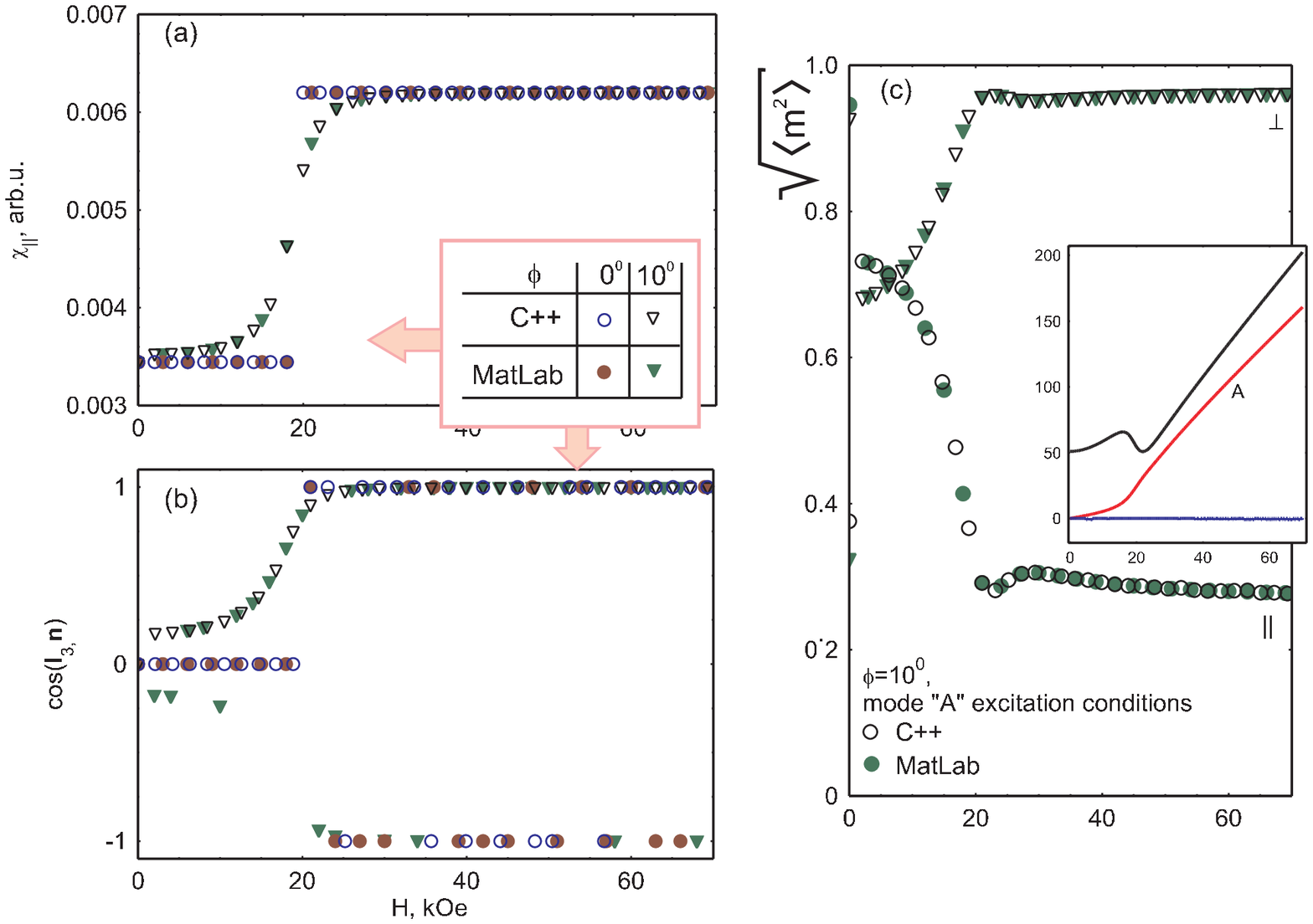, width=\figwidth, clip=}
  \caption{Comparison of modeled static properties (panels (a) and (b)) and excitation conditions for one of the AFMR mode (panel(c)) calculated for the test case of \csnicl{} by different implementations of numeric procedure. (a) --- longitudinal static susceptibility; (b) --- orientation of spin structure with respect to the magnetic field; (c) --- excitation conditions for the mode denoted as ``A'' on the insert (triangles correspond to conventional transverse pumping, circles --- to the longitudinal pumping).}\label{fig:csnicl-compare}
\end{figure}

Both implementations of numeric procedure reproduce analytical $f(H)$ curves well (Figs.\ref{fig:csnicl-cpp} and \ref{fig:csnicl-matlab}). Analytical $f(H)$ dependences can be calculated for \csnicl{} in arbitrary field orientation \cite{csnicl-afmr}, we performed our modeling in certain representative cases: $\vec{H}||z$ ($\phi=0^\circ$), slightly canted field ($\phi=10^\circ$) and $\vec{H}\perp z$ ($\phi=90^\circ$).

C++ implementation demonstrated numeric instability at $\vec{H}||z$ for $H<H_c$. At these fields zero frequency mode is two-fold degenerated and numeric uncertainties of calculation lead to imaginary roots for $\omega^2$ in $det{\cal M}=0$ equation (two fold degeneracy of cubic equation root means that cubical parabola is tangent to $y=0$ at some point, condition extremely sensitive to coefficients definition). As described above, this error was handled by setting output frequency to dummy value of $-1$. Since degeneracy of oscillation modes is a rare event, we believe that this do not cause big discomfort. MatLab implementation was free from this problem.

Static properties and excitation conditions modeled by both implementations coincide (Fig.\ref{fig:csnicl-compare}). \csnicl{} demonstrate spin-reorientation transition at $H_c\approx 19$kOe, at this field normal to the plane of the spin structure rotates along the field direction. This is reproduced by both approaches. As the field is canted from the symmetry axis ($\phi=10^\circ$ case) spin-reorientation became smeared over certain field range, as expected. Note that when looking for the global minimum numeric procedure randomly switches between equivalent orientations $\vec{l}_3||\vec{H}$ and $\vec{l}_3||-\vec{H}$ above $H_c$. As these orientations are equivalent, this does not cause any problem when calculating physically observed quantities (oscillation frequencies, susceptibilities etc.). This issue can be evaded by setting on local minimum search option in C++ implementation, which inherits starting approximation for energy minimum search from previous field point. Excitation conditions are shown for one of the modes for $\phi=10^\circ$ case. Calculation coincide for both implementations, note that average longitudinal magnetization magnitude is not negligible for this mode.

\clearpage

\subsection{Case of \AlMn{}}
\begin{figure}
  \centering
 \epsfig{file=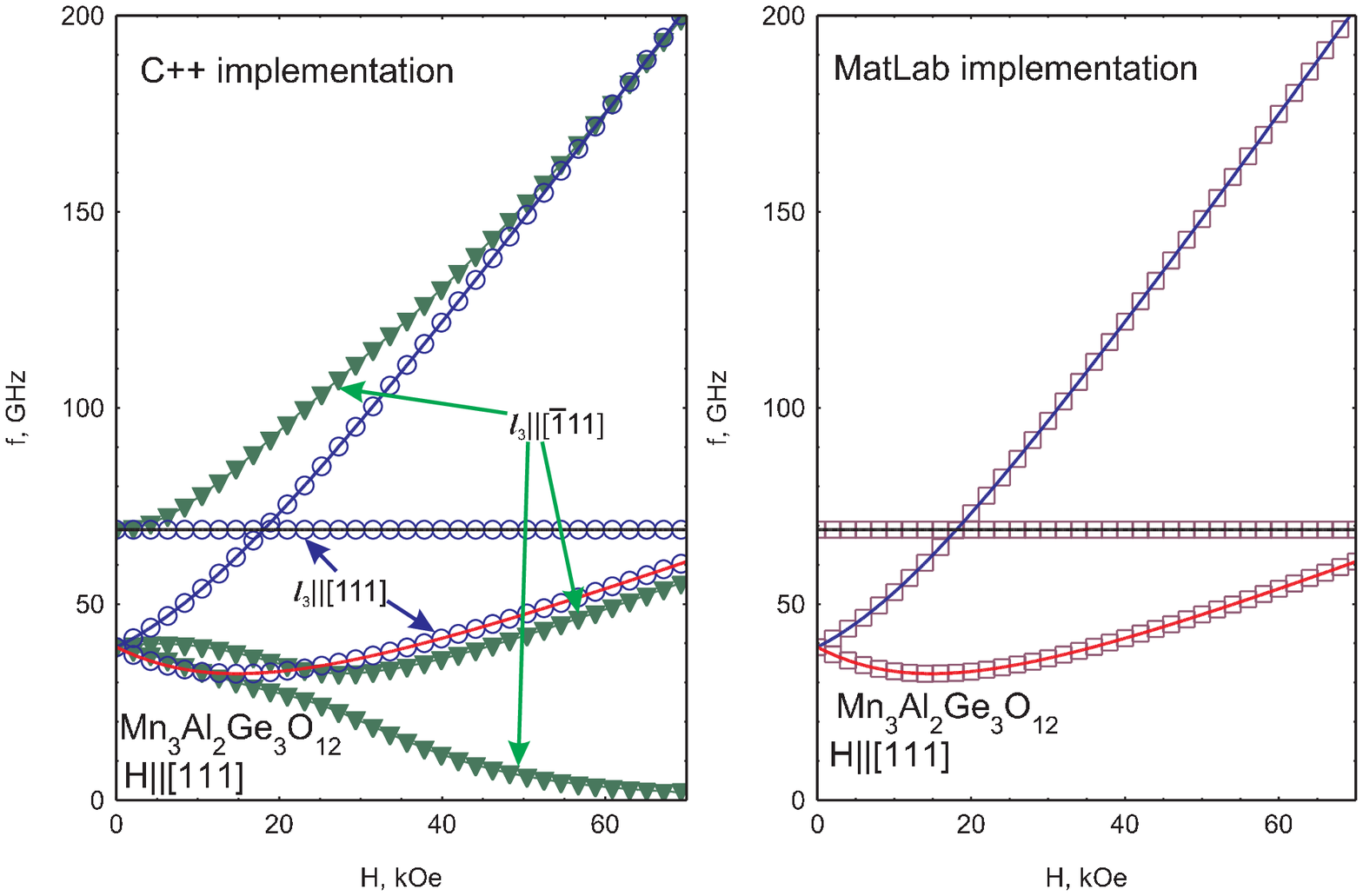, width=\figwidth, clip=}
  \caption{Modeled $f(H)$ curves for \AlMn{} at $\vec{H}||[111]$. Left panel --- C++ implementation, right panel --- MatLab implementation. Thick solid lines on both panels are analytical calculations. On the left panel (C++ implementation) $f(H)$ curves for stable (circles) and metastable (triangles) domains are calculated. }\label{fig:almn111}
\end{figure}

\begin{figure}
  \centering
 \epsfig{file=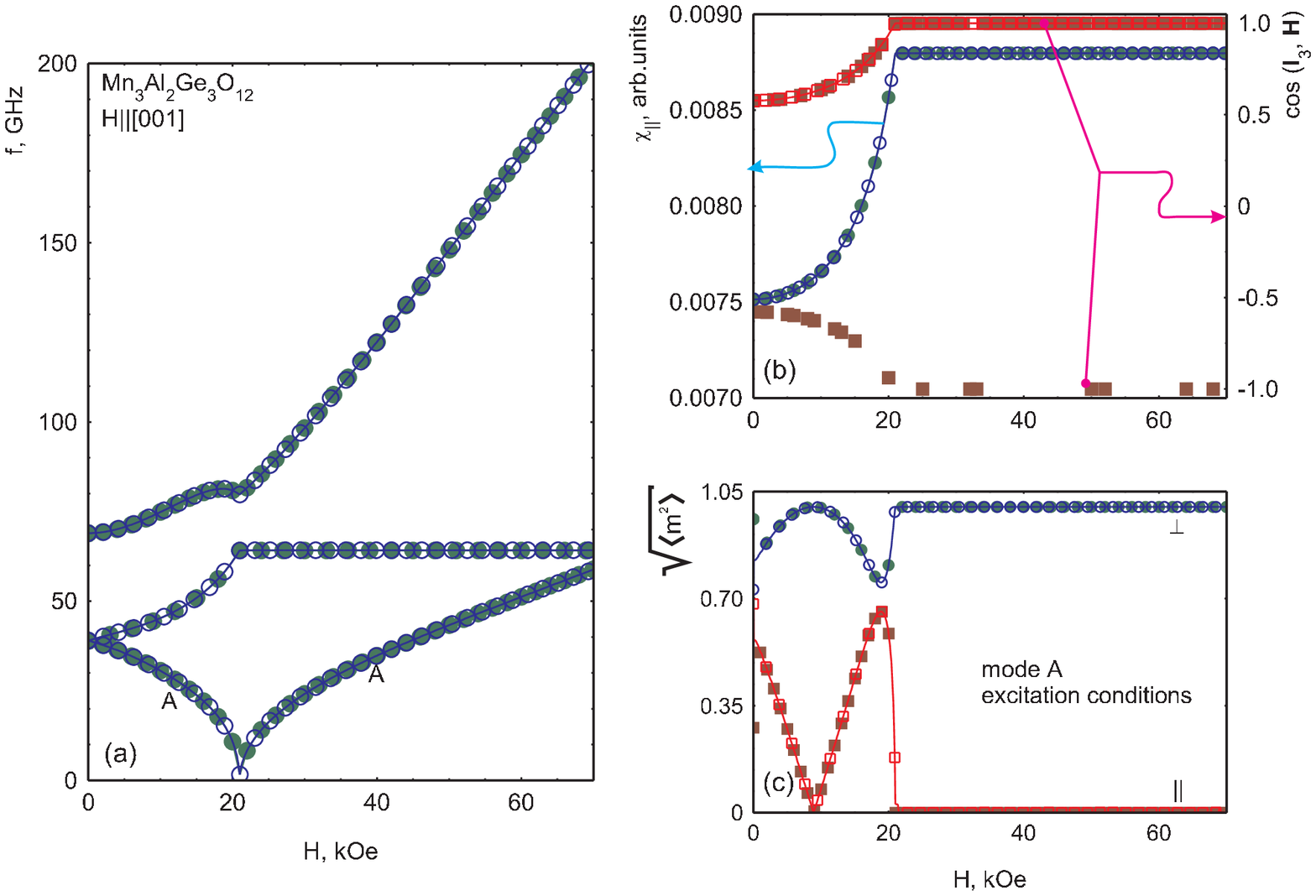, width=\figwidth, clip=}
  \caption{Modeled $f(H)$ curves, static properties and excitation conditions for \AlMn{} at $\vec{H}||[001]$. Open symbols --- C++ implementation, closed symbols --- MatLab implementation, curves --- guide to the eye. (a) $f(H)$ dependence, (b) longitudinal susceptibility (circles, left Y-axis) and orientation of the normal to the plane of the spin structure with respect to the field (squares, right Y-axis), (c) average oscillating transverse (circles) and longitudinal (squares) magnetization for mode A (see panel (a)). }\label{fig:almn001}
\end{figure}

\begin{figure}
  \centering
 \epsfig{file=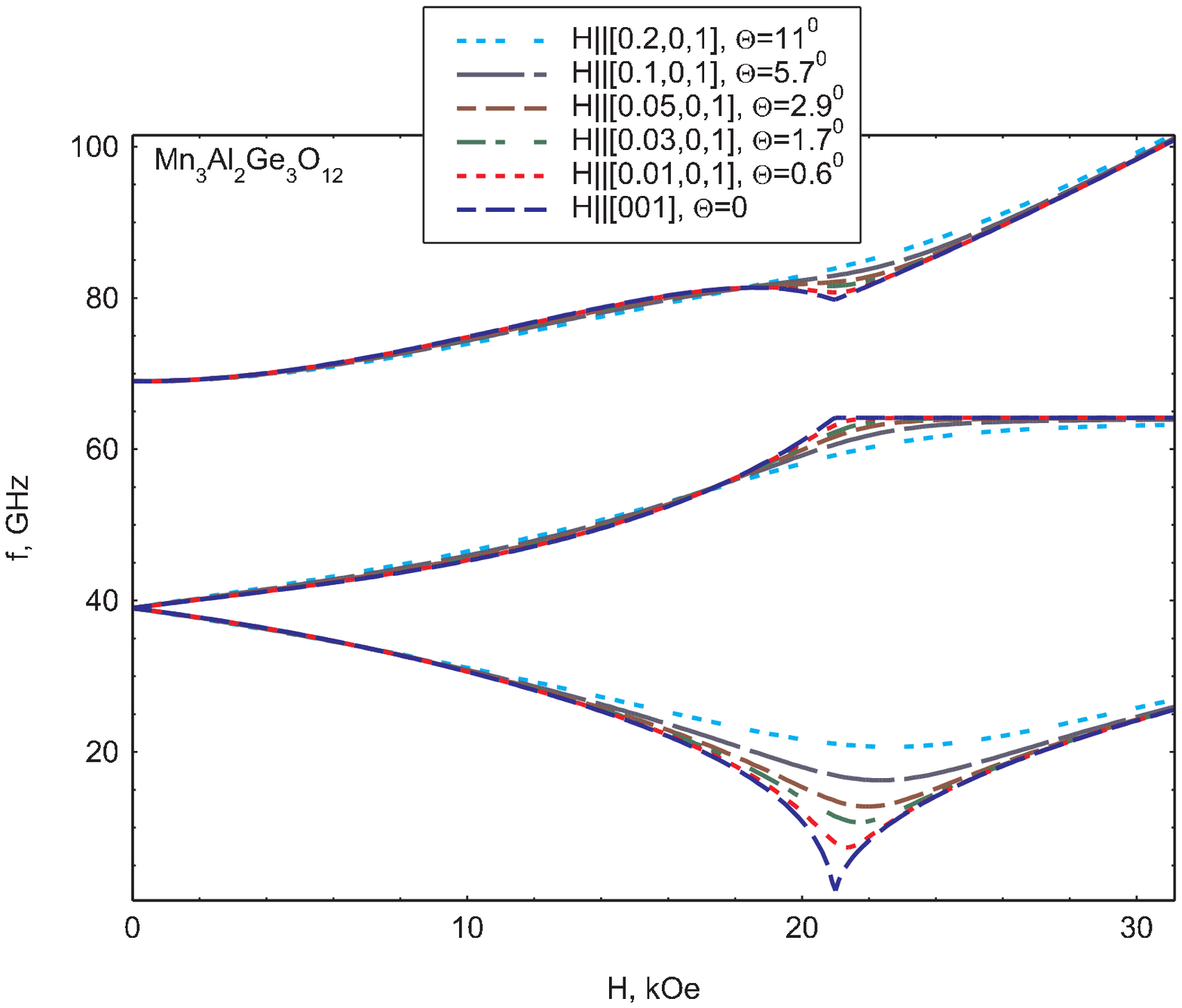, width=\figwidth, clip=}
  \caption{Modeled $f(H)$ curves for \AlMn{} at the field $\vec{H}||[\delta 01]$. C++ implementation used only.}\label{fig:almn-canted}
\end{figure}

In the case of \AlMn{} at zero field four equivalent magnetic domains are possible at zero field, planes of the spin structure in these domains lie orthogonal to different $\langle 111 \rangle$ axes of the cubic crystal. For the field applied along $[111]$ axis one of these domains is stable, while other remains metastable. Dynamics equations can be obtained analytically for the stable domain, we  have found than both implementation yields the same numeric results (Fig.\ref{fig:almn111}). Additionally, one can obtain $f(H)$ curve for metastable domain making use of local minimum search option in C++ implementation.

At $\vec{H}||[001]$ all domains are equivalent. In this orientation spin reorientation takes place: planes of the spin structure begin to rotate as the field is applied and complete reorientation by setting  plane of the spin structure orthogonal to the applied field at the critical field $H_c\approx 21$kOe. No analytical solution of dynamics equation is possible in this orientation, numeric methods easily solve this problem (Fig.\ref{fig:almn001}). Both implementation results coincide. Again, see panel (b) of Fig.\ref{fig:almn001}, numeric minimum energy search procedure sometimes switches between equivalent domains, but this does not affect observable quantities ($f(H)$ or $\chi(H)$).

As it is well known, spin reorientation is very sensitive to the exact orientation of the magnetic field with respect to the crystallographic axis. Numeric methods allow to model this situation as well (Fig.\ref{fig:almn-canted}), which eases analysis of experimental data, allowing to estimate canting of the sample, for example.

\clearpage

\subsection{Case of \licuo{}}
\begin{figure}
  \centering
\epsfig{file=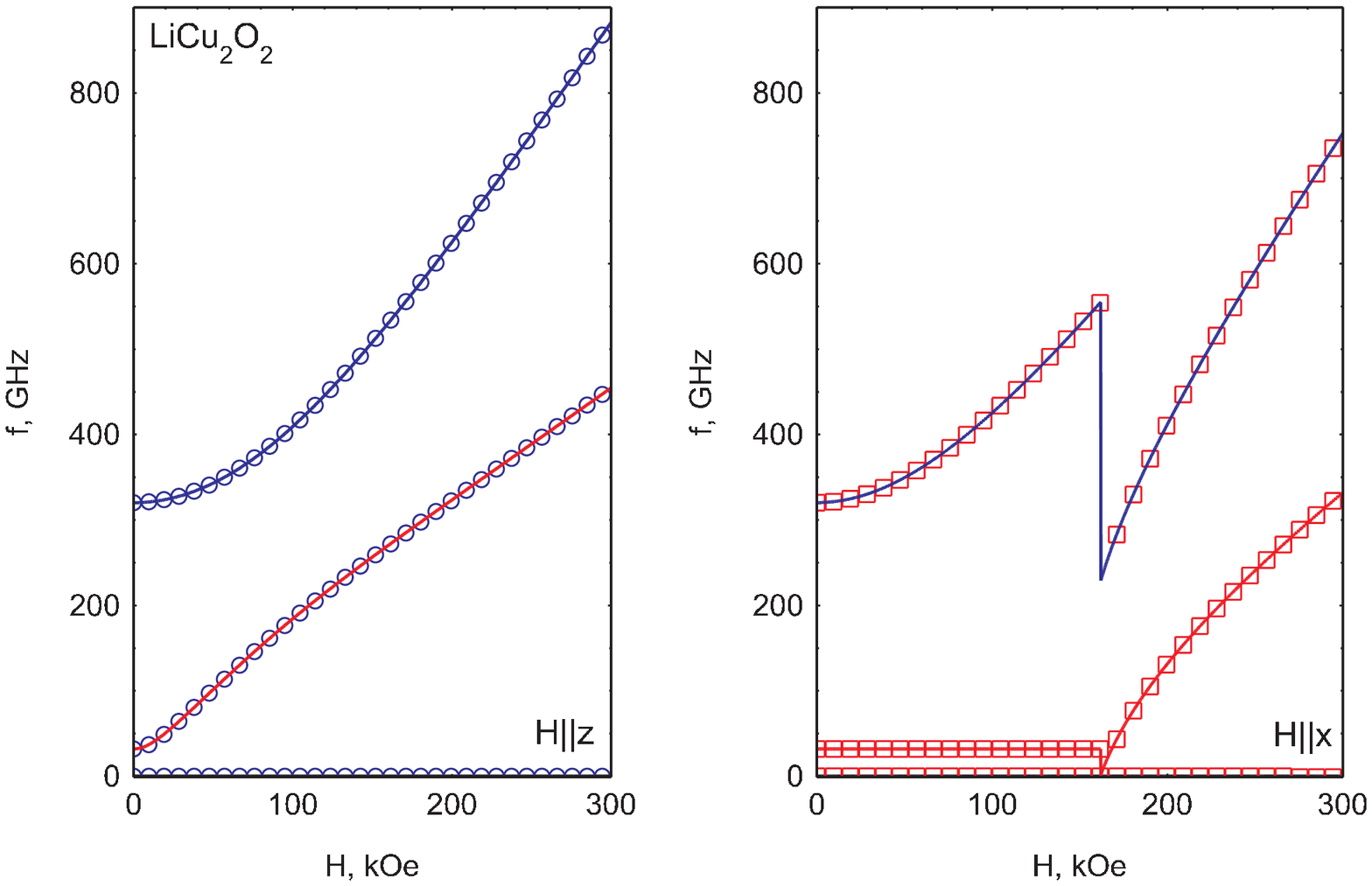, width=\figwidth, clip=}
  \caption{Modeled $f(H)$ curves for \licuo{}, C++ implementation of the numeric procedure. Left panel $\vec{H}||z$, right panel $\vec{H}||x$. Thick solid lines on both panels are analytical calculations.}\label{fig:licuo-cpp}
\end{figure}

\begin{figure}
  \centering
\epsfig{file=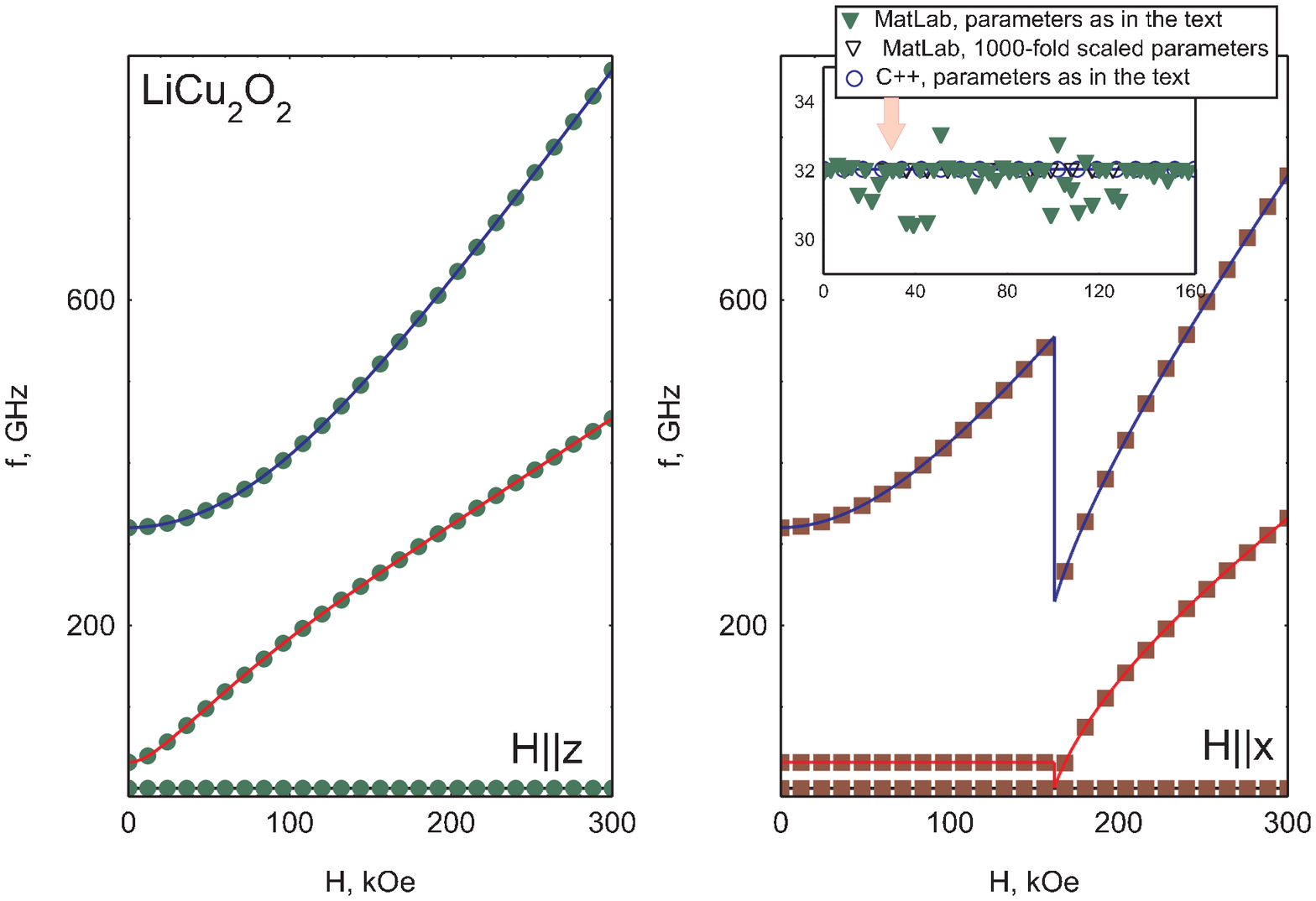, width=\figwidth, clip=}
  \caption{Modeled $f(H)$ curves for \licuo{}, MatLab implementation of the numeric procedure. Left panel $\vec{H}||z$, right panel $\vec{H}||x$. Thick solid lines on both panels are analytical calculations. Insert on the left panel illustrates numeric uncertainties of the 32 GHz AFMR mode calculation below $H_c$.}\label{fig:licuo-matlab}
\end{figure}

\begin{figure}
  \centering
\epsfig{file=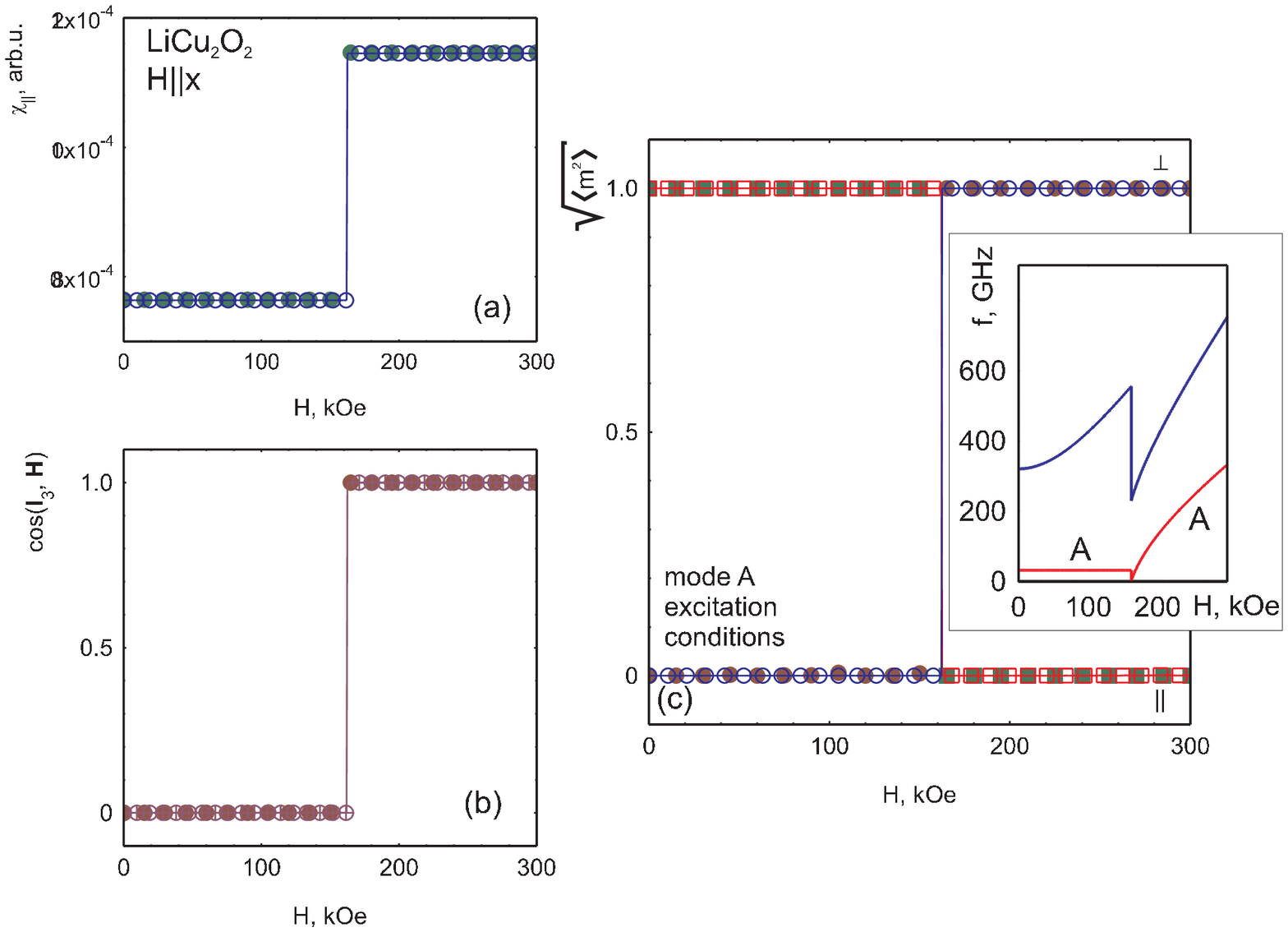, width=\figwidth, clip=}
  \caption{Modeled static properties and excitation conditions for one of the AFMR modes for \licuo{}, $\vec{H}||x$. Open symbols --- C++ implementation, closed symbols --- MatLab implementation, curves are guides to the eye. (a) longitudinal susceptibility, (b) orientation of the spin plane with respect to the magnetic field, (c) average amplitude of longitudinal (squares) and transverse (circles) components of oscillating magnetization in AFMR mode ``A'' (see inset). }\label{fig:licuo-comp}
\end{figure}

AFMR modes for \licuo{} can be found analytically in main orientations of applied field. Results of numeric procedure (Figs.\ref{fig:licuo-cpp} and \ref{fig:licuo-matlab}) fits analytical curves well. As anisotropy constants for \licuo{} are very close (they differ by 1\%) accuracy of the equilibrium position determination affects strongly one of the AFMR modes at $\vec{H}||x$: the field independent $f\approx32$GHz mode corresponds to the oscillation of the spin structure in the $(yz)$ plane and its frequency is determined by difference of anisotropy constants. We have found, that MatLab implementation is more sensitive to this issue (see inset at the Fig.\ref{fig:licuo-matlab}) with uncertainties up to 1 GHz (3\% accuracy) for the set of parameters specified in this text (section \ref{sec:defs}). We have found that stability of the numeric output can be improved by scaling model parameters (anisotropy constants and $I_i$ constants) by the factor of 1000 (as it is described in the main paper, scaling factor can be chosen arbitrary). It seems that this issue is due to some built-in rounding restrictions in MatLab, C++ implementation was free of this issue.

As for other examples, we model static properties and excitation conditions for one of AFMR modes (Fig.\ref{fig:licuo-comp}). Both implementations results coincide. At $\vec{H}||x$ a sudden spin-reorientation is expected for \licuo{}, at this field spin plane rotates normally to the applied field. Note that excitation conditions for the AFMR mode, which is field-independent below $H_c$ also suddenly change at spin-reorientation: the oscillating magnetization is parallel to the applied field below $H_c$ and orthogonal to the applied field above $H_c$.

\clearpage

\section{Conclusions (supplementary material)}
We have checked C++ and MatLab implementations of numerical algorithm for description of AFMR $f(H)$ dependences in noncollinear antiferromagnets against some known analytically solvable cases. We have found that modeled results fits to analytical results well, both implementation results coincide. We have not found serious instabilities in the algorithm implementations, several minor issues related to strong degeneracy of the particular cases were observed and discussed.


\begin{thebibliography}{}
\bibitem{andrmar} A.F. Andreev, V.I. Marchenko, Sov. Phys. Usp. \textbf{130}, 39
(1980)

\bibitem{csnicl-neutrons}     V.J. Minkiewicz, D.E. Cox, G. Shirane, Solid State Communications \textbf{8}, 1001 (1970)

\bibitem{csnicl-afmr} I.A. Zaliznyak, V.I. Marchenko, S.V. Petrov, L.A. Prozorova, A.V. Chubukov, JETP Letters \textbf{47}, 211 (1988)

\bibitem{garnet-prandl} W. Prandl, Physica Status Solidi (b) \textbf{55}, K159 (1973)

\bibitem{garnet-afmr} L.A. Prozorova, V.I. Marchenko, Yu.V. Krasnyak,  JETP Letters, \textbf{41}, 637 (1985)

\bibitem{licuo-nmr} A.A. Gippius,   E.N. Morozova, A.S. Moskvin, A.V. Zalessky, A.A. Bush, M. Baenitz, H. Rosner, and S.L. Drechsler Phys. Rev. B \textbf{70}, 020406(R) (2004)

\bibitem{licuo-neutrons} T.  Masuda,  A.  Zheludev,  A.  Bush,  M.  Markina,  and
A. Vasiliev, Phys. Rev. Lett. \textbf{92}, 177201 (2004)

\bibitem{svistov-licuo} L.E. Svistov, L.A. Prozorova, A.M. Farutin, A.A. Gippius, K.S.
 Okhotnikov, A.A. Bush, K.E. Kamentsev, E. A. Tishchenko, JETP \textbf{108}, 1000 (2009)

\bibitem{stewart} J.R. Stewart, G. Ehlers, A.S. Wills, S.T. Bramwell and J.S. Gardner
Journal of Physics: Condensed Matter, \textbf{16}, L321 (2004)

\bibitem{nagomiya} T. Nagamiya, K. Yosida and R. Kubo, Advances in Physics, \textbf{4}, 1 (1955)

\bibitem{tanaka} H. Tanaka, S. Teraoka, E. Kakehashi, K. Iio, and K. Nagata, J. Phys. Soc. Jpn. \textbf{57},  3979 (1988)

\bibitem{chubukov} A.V.Chubukov and D.I.Golosov, J.Phys.:Condens.Matter, \textbf{3}, 69 (1991)

\bibitem{sosin2009} S.S.Sosin, L.A.Prozorova, P. Bonville, M.E.Zhitomirsky, Phys. Rev. B, \textbf{79}, 014419 (2009)

\bibitem{spinW} SpinW Homepage by S.T\'{o}th, https://www.psi.ch/spinw/spinw

\bibitem{home} Authors Web-page, http://www.kapitza.ras.ru/rgroups/esrgroup/, see ``NuMA: Numeric Methods for Antiferromagnets'' section of the web-page

\bibitem{sosin2006} S.S.Sosin, A.I.Smirnov, L.A.Prozorova, G.Balakrishnan, M.E.Zhitomirsky, Phys.Rev B, \textbf{73}, 212402 (2006)

\bibitem{mar1999} V.I.Marchenko and A.M.Tikhonov, JETP Letters \textbf{69}, 44 (1999)

\bibitem{vasiliev2001} A.N.Vasil'ev, V.I.Marchenko, A.I.Smirnov,  S.S.Sosin, H.Yamada and Y.Ueda, Phys. Rev. B \textbf{64}, 174403 (2001)

\bibitem{far-glaz} V.N.Glazkov, A.M.Farutin, V.Tsurkan, H-A. Krug von Nidda and A.Loidl, Phys. Rev. B \textbf{79}, 024431 (2009)

\bibitem{farmar-HF} A.M.Farutin, V.I.Marchenko, JETP Lett. \textbf{83}, 238 (2006)

\bibitem{NR} W.H.Press, S.A.Teukolsky, W.T.Vetterling, B.P.Flannery, Numerical Recipes: The Art of Scientific Computing, Cambridge University Press (2007), http://numerical.recipes

\bibitem{udalov} O.G.Udalov, JETP \textbf{113},  490 (2011)

\end{thebibliography}
\end{document}